\documentclass[twocolumn,aps,prb,floatfix,showpacs]{revtex4}
\usepackage{graphicx,graphics,color,epsfig}
\usepackage{amsmath}
\usepackage{amssymb}

\begin{document}
\def\runtitle{Spin-lattice model of RbCoBr$_3$}
\def\runauthor{Tota {Nakamura} and Yoichi {Nishiwaki}}

\title{%
Spin-lattice model of Magneto-electric Transitions in RbCoBr$_3$
}

\author{Tota Nakamura$^1$ and Yoichi Nishiwaki$^2$ }

\affiliation{%
$^1$
Faculty of Engineering, Shibaura Institute of Technology,
         Minuma-ku, Saitama  330-8570, Japan
\\
$^2$
Department of Physics, Tokyo Women's Medical University, 
         Shinjuku-ku, Tokyo, 162-8666, Japan
}

\date{\today}

\begin{abstract}
Extensive Monte Carlo simulations are performed to analyze
a recent neutron diffraction experiment
on a distorted triangular lattice compound RbCoBr$_3$.
We consider a spin-lattice model, where both spin and lattice are
Ising variables.
This model explains well successive magnetic and dielectric transitions
observed in the experiment.
The exchange interaction parameters and the spin-lattice coupling
are estimated.
It is found that the spin-lattice coupling is important to explain
the slow growth of a ferrimagnetic order.
The present simulations were made possible
by developing a new Monte Carlo algorithm, which accelerates slow
Monte Carlo dynamics
of quasi-one-dimensional frustrated systems.
\end{abstract}

\maketitle

\section{Introduction}

Frustrated magnets have been attracting much 
interest for many decades.\cite{HFM2006}
The ordinary magnetic order is destroyed and the ground state 
may remain disordered or turn into an exotic state.
The ground state of a frustrated system is usually unstable against
a small perturbation.
The system manages to find a way to relax frustration and change the state.
We may design and control material functions using the frustration effects.

The ABX$_3$-type compounds are well-known frustrated magnets.
The lattice structure is the stacked triangular lattice. 
There is frustration
when the nearest-neighbor interactions are antiferromagnetic.
Successive magnetic phase transitions occur because of the strong 
frustration when the spins have the Ising anisotropy.\cite{CsCoBr3,CsCoCl3}
The low-temperature magnetic structure is the ferrimagnetic state.
There exists a partially-disordered (PD) phase between the paramagnetic
phase and the ferrimagnetic phase.
In the PD phase, one of the three sublattices is completely disordered, while
the other two sublattices take antiferromagnetic configurations.
There is no structural phase transition in most compounds.
The system remains fully frustrated down to the lowest temperature.

The KNiCl$_3$-family compounds are exceptional in that they
exhibit structural phase transitions.\cite{visser,ex55,RbFeBr3-experi}
We can observe the structural phase transitions 
by the dielectric measurements because each BX$_3$ chain
has a negative charge.
These compounds have both magnetic and dielectric characteristics:
we call them the magneto-dielectric compounds.
The magnetic phase transitions and the structural (dielectric) phase 
transitions usually occur at different temperatures.
However, 
Morishita {\it et al.} found that both transitions
occur at the same temperature in 
RbCoBr$_3$.\cite{morishita52,morishitaC}
It is a very rare case among the KNiCl$_3$-family compounds.
Magnetic and dielectric measurements\cite{nishiwaki2,experi-mag}
found that the phase transitions in RbCoBr$_3$ are quite unusual
in the following points compared with other compounds:
\begin{itemize}
\item[(i)]
The dielectric transition temperature 37~K of RbCoBr$_3$
is very low compared with other compounds, for which the transition takes place
around the room temperatures.
The energy scale of the structural (dielectric) system in RbCoBr$_3$
seems to be suppressed somehow.
\item[(ii)]
The temperature dependence of the dielectric constant does not exhibit
a diverging behavior.
This is clearly different from another KNiCl$_3$-family compound RbFeBr$_3$, 
which exhibits sharp divergence at the transition
temperature, 34.4 K.\cite{RbFeBr3-experi}
\item[(iii)]
The increase of the spontaneous polarization below the dielectric 
transition temperature is very slow, while that of RbFeBr$_3$ is 
very sharp.
\item[(iv)]
The magnetic PD phase appears in a very narrow temperature region.
The first neutron measurement suggested that it might disappear.\cite{nishiwaki2}
A recent improved neutron experiment \cite{experi-f} made it clear that 
it exists between 31 K and 37 K.
This is also a clear difference from other compounds such as
CsCoBr$_3$\cite{CsCoBr3} and CsCoCl$_3$.\cite{CsCoCl3}
\item[(v)]
The growth of the ferrimagnetic order is very slow. 
The neutron count increases linearly with the temperature
decrease in the low-temperature phase.
\end{itemize}
These characteristic behaviors suggest that there is an unknown
mechanism of interplay between the magnetic system and the dielectric system 
in RbCoBr$_3$.

An aim of this paper is to propose a proper theoretical model that
{\it quantitatively} explains the experimental results of RbCoBr$_3$.
A well-known theoretical model for the ABX$_3$ compounds is the 
antiferromagnetic spin system on the stacked triangular
lattice.\cite{shiba,matsubara-ina,kurata,koseki,todoroki}
However, the ordinary spin model without a coupling to the dielectric system
is not sufficient to explain RbCoBr$_3$.
The chain-mean-field theory\cite{shiba} gives the magnitude of the 
second-nearest-neighbor magnetic interactions on the $c$-plain 
$|J_2|\simeq 1$ K, which is comparable to that of the nearest-neighbor 
interactions $|J_1|\simeq  2.5$ K.\cite{morishitaC}
This is not acceptable from the experimental point of view.

In the present paper, we use the spin-lattice model
proposed by Shirahata and Nakamura.\cite{shiramaster}
This model showed that the PD phase may disappear because of
the relaxation of frustration by the lattice distortion;
each of the spin system and the lattice system relaxes frustration of the other.
A single transition may occur from a paramagnetic and paraelectric phase to
the ground state phase without experiencing the intermediate PD phase.
Shirahata and Nakamura also noticed that the cooperation 
between the spin system and the lattice system
works only when the energy
scale of the lattice system is comparable with that of the spin system.

In this paper, we refine the above spin-lattice model.
We find that a soft lattice system coupled with a spin system
explains the interesting material RbCoBr$_3$.
We thereby clarify
the origin of the characteristic behaviors of this compound.
For the purpose,
we develop a new Monte Carlo (MC) algorithm, which eliminates slow MC dynamics
in quasi-one-dimensional frustrated spin systems.
We have performed extensive MC simulations and determined various physical
parameters.

We explain our model Hamiltonian in Sec. \ref{sec:model}.
A numerical method is explained in Sec. \ref{sec:method}, and the results
are presented in Sec. \ref{sec:results}.
Discussions are given in Sec. \ref{sec:discussion}.

\section{Theoretical Model}
\label{sec:model}

\subsection{Structure of ABX$_3$ compounds}

The lattice structure of ABX$_3$-type compounds is 
the stacked triangular lattice.
Face-sharing BX$_6$ octahedra run along the $c$-axis forming a BX$_3$ chain.
Magnetic B$^{2+}$ ions form an equilateral triangular lattice on the
$c$-plane, which causes frustration.
Exchange interactions along the BX$_3$ chains, $J_c$, are antiferromagnetic.
The magnitude of $J_c$ is much larger than that of the nearest-neighbor 
interactions $J_1$ on the c-plane: $|J_c|\gg |J_1|$.
Therefore, 
this spin system can be considered as
a quasi-one-dimensional system with frustration on the $c$-plane.

\begin{figure}
\includegraphics[width=2.6cm]{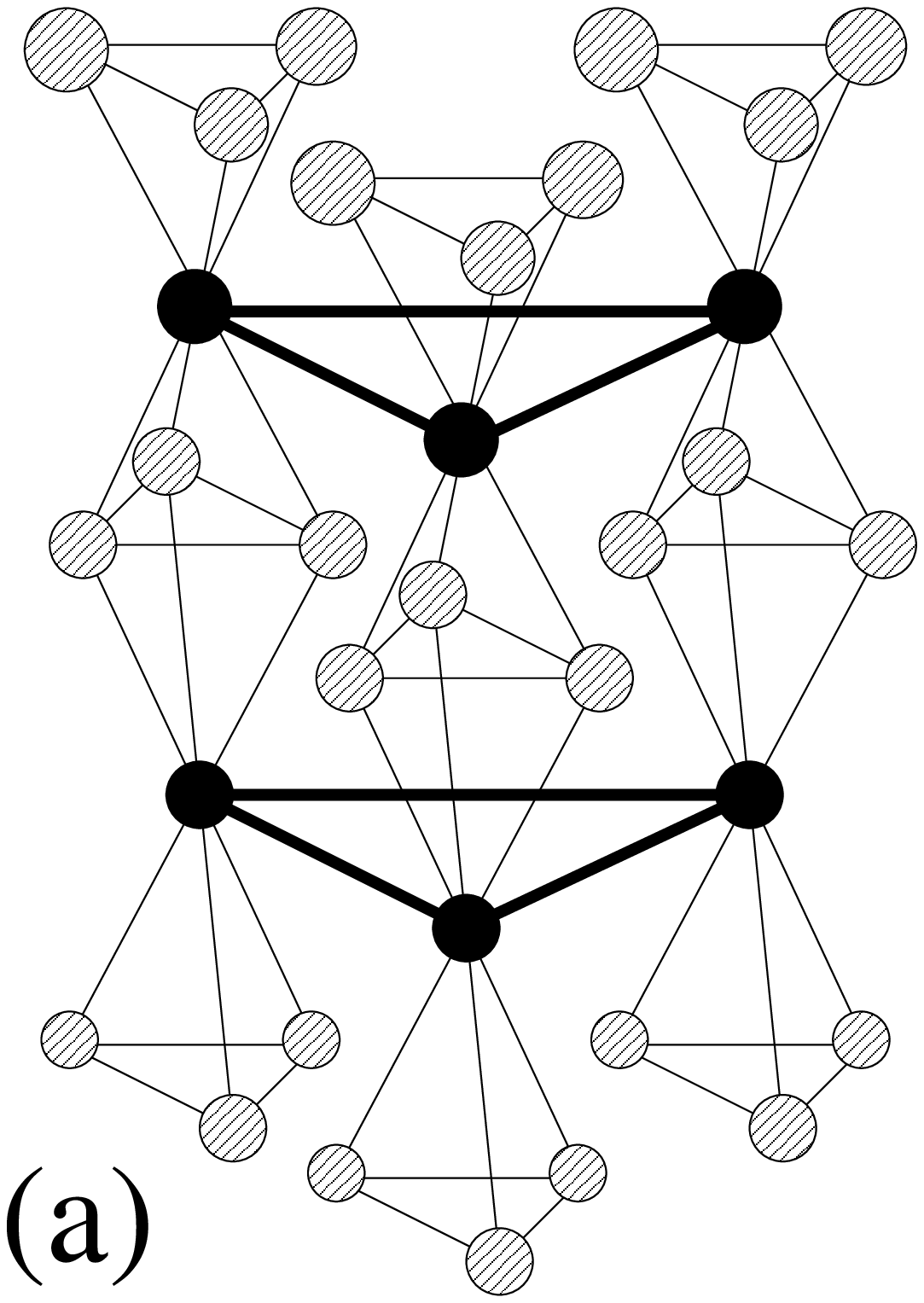}
\hfil
\includegraphics[width=2.6cm]{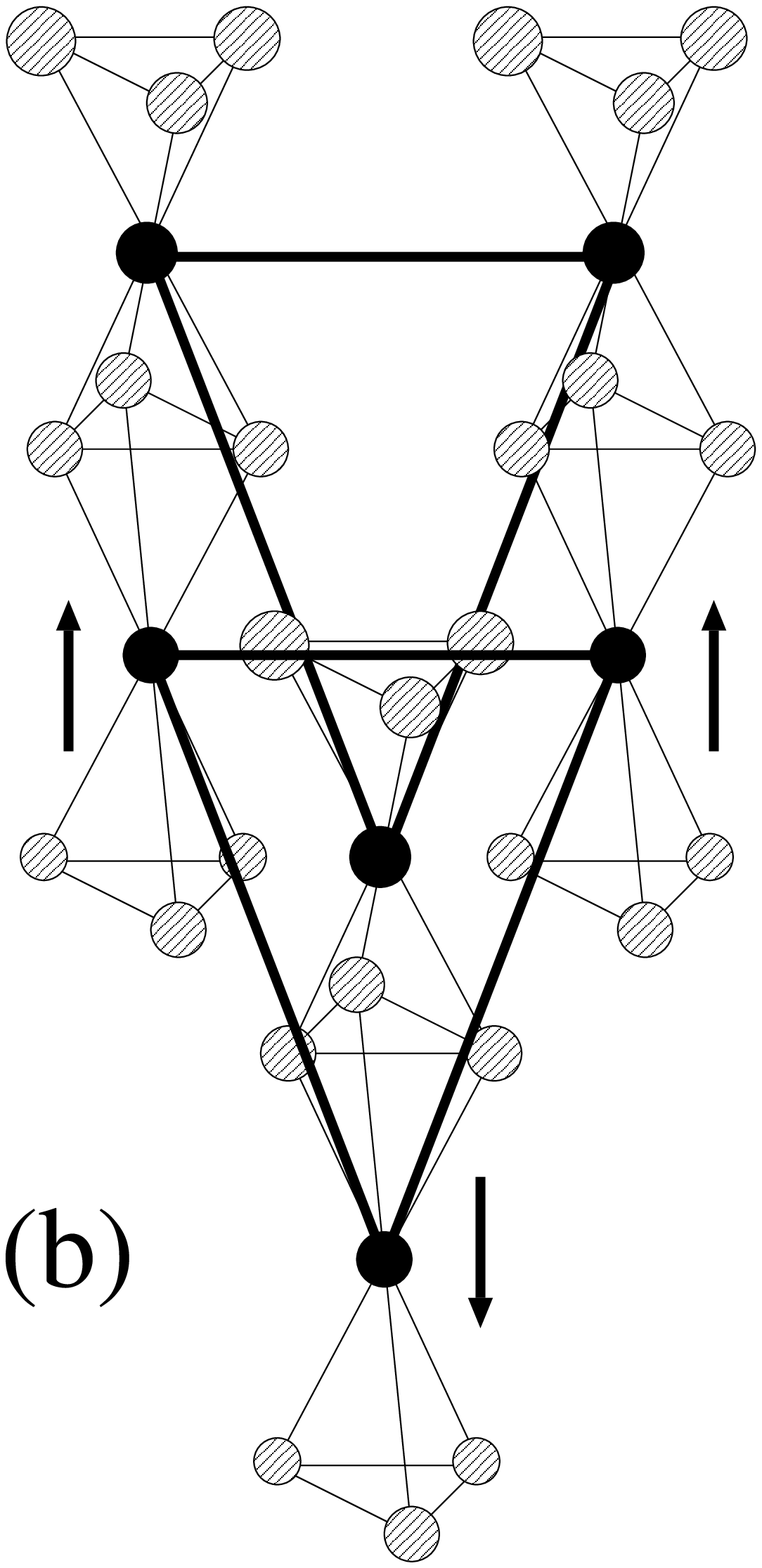}

\includegraphics[width=2.6cm]{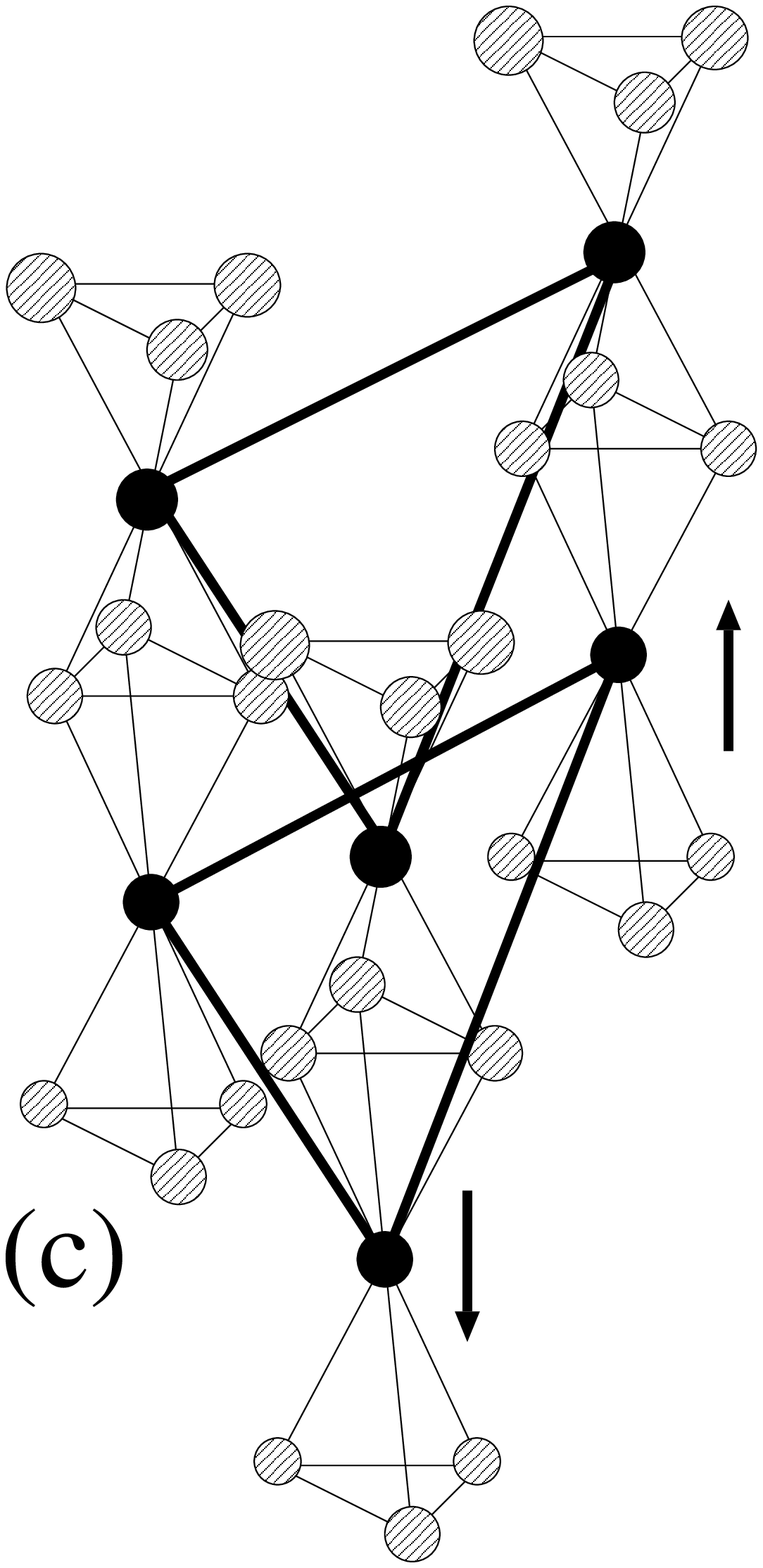}
\hfil
\includegraphics[width=2.6cm]{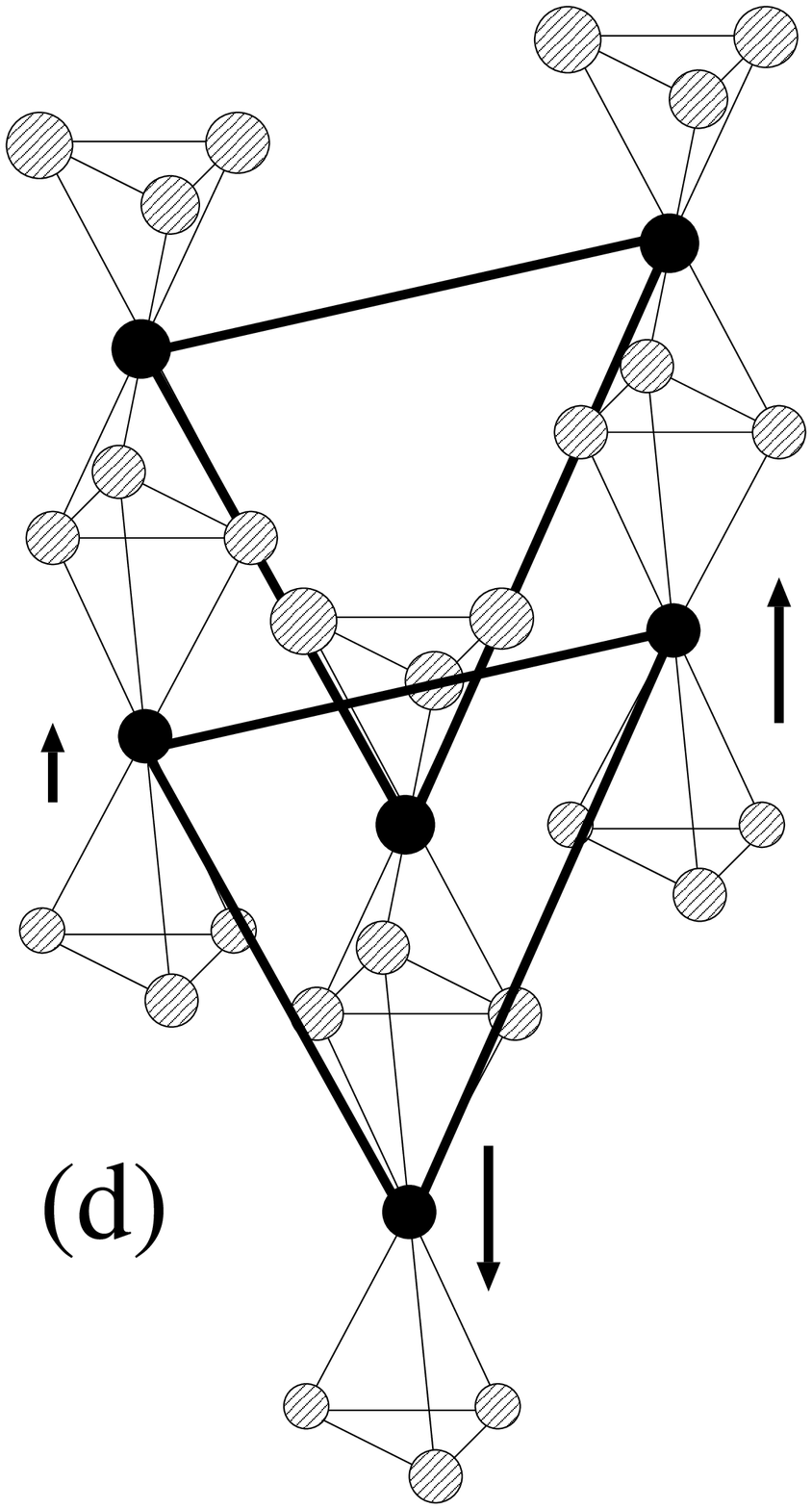}
\caption{
Typical crystal structures of ABX$_3$-type compounds.
A-ions are omitted.
Black circles depict magnetic B$^{2+}$ ions and grey circles depict X$^-$ ions.
Each arrow depicts the shift direction of a chain.
(a) A symmetric structure at high temperatures. The space group is $P6_3/mmc$.
(b) A room-temperature KNiCl$_3$ structure. The space group is $P6_3cm$.
We call this structure ``lattice-Ferri".
(c) A low-temperature structure. The space group is $P\bar{3}c1$.
We call this structure ``lattice-PD".
(d) Another low-temperature structure. The space group is $P3c1$.
We call this structure a ``three-sublattice lattice-Ferri''.
Each sublattice polarization takes a different value.
}
\label{fig:lattice}
\end{figure}

The typical lattice structure of ABX$_3$-type compounds
at high temperatures is shown in Fig.~\ref{fig:lattice} (a).
The space group is $P6_3/mmc$.
Magnetic ions forming an equilateral triangular lattice
sit on a level plane.
This structure remains down to the lowest temperature in most compounds.
In the KNiCl$_3$ family, the structural phase transitions occur
as we decrease the temperature.
Each BX$_3$ chain shifts upward or downward keeping the relative distance 
between the atoms.
One of the lattice structures after the structural phase transitions 
is shown in Fig.~\ref{fig:lattice} (b), where
two sublattices on the triangular lattice shift upward with the same amount
while one sublattice shifts downward.
The space group is $P6_3cm$.
It is the ferrielectric structure of KNiCl$_3$ observed at the 
room temperatures.\cite{visser}
We refer to this structure as ``lattice-Ferri" or 
``$\uparrow$-$\uparrow$-$\downarrow$" in this paper.
Another possible structure is a configuration with one sublattice shifting
upward, one shifting downward and the third unchanged as shown in 
Fig.~\ref{fig:lattice} (c).
The space  group is $P\bar{3}c1$.
We refer to this structure as ``lattice-PD" or ``$\uparrow$-$\downarrow$-0" 
in this paper.
Nishiwaki and Todoroki\cite{nishi-todo.MF} discussed the appearance
of the three-sublattice ferrielectric state  in RbCoBr$_3$ 
using the mean-field approximation.
It is a structure with the $\uparrow$-$\uparrow$-$\downarrow$ configuration
but the amount of displacement in each sublattice is different from the others,
as shown in Fig.~\ref{fig:lattice} (d).
The space group is $P3c1$.
We refer to this structure as ``three-sublattice lattice-Ferri" in order
to distinguish from the ``lattice-Ferri" structure
of Fig.~\ref{fig:lattice}(b), which has the two-sublattice order.

\subsection{Spin-lattice model}

We consider a model 
on the stacked triangular lattice
with spin and lattice degrees of freedom.\cite{shiramaster}
The size in the $a$ and $b$ directions is $L$,
while the size in the $c$ direction is $L_c$.
There are a spin variable $S_{ij}$ and a lattice 
variable $\sigma_{ij}$ at each site.
Here, the subscript $i$ denotes the position in the $c$-axis, while $j$ denotes
the position on the $c$-plane.
We define each spin as an Ising variable
$S_{ij}=\pm 1/2$ because Co ions have the Ising anisotropy.

The lattice variable $\sigma_{ij}$ denotes the displacement
from the symmetric lattice point along the $c$-axis.
We approximate the displacement with the Ising variable 
as $\sigma_{ij}=\pm 1/2$;
each ion shifts either upward ($\sigma_{ij}=1/2$) or 
downward ($\sigma_{ij}=-1/2$).
The reason of approximating the present lattice system with
the Ising variables is as follows.
The symmetric lattice structure with zero displacement
as shown in Fig.~\ref{fig:lattice}(a) ($P6_3/mmc$)
appears at high temperatures.
Structural phase transitions occur successively as the temperature decreases.
The lattice configuration is 
the $\uparrow$-$\downarrow$-0 state (Fig.~\ref{fig:lattice}(c)) in the 
intermediate phase and is
the $\uparrow$-$\uparrow$-$\downarrow$ state (Fig.~\ref{fig:lattice}(b))
in the low-temperature phase.
This is analogous to the successive magnetic phase transitions 
from the paramagnetic phase to the PD phase and to the ferrimagnetic phase
in the {\it Ising} model on the stacked triangular lattice.
In the previous paper,\cite{shiramaster} 
we considered a lattice variable taking 
three states, $+1, 0, -1$.
Here, we omit a state $\sigma_{ij}=0$.
The $\sigma_{ij}=0$ state can be represented by
a mixture of the $\sigma_{ij}=1/2$ state and the $\sigma_{ij}=-1/2$ state.
A chain shift is the sum of the ion displacements along the chain
in the experiment.

The Hamiltonian consists of the lattice part $\mathcal{H}_{\rm L}$ and
the spin part $\mathcal{H}_{\rm S}$:
\begin{equation}
\mathcal{H}=\mathcal{H}_\mathrm{L} + \mathcal{H}_\mathrm{S},
\end  {equation}
where
\begin{eqnarray}
\mathcal{H}_\mathrm{L}=
&-&
2J_c^\mathrm{L} 
 \sum_{i,j}
\sigma_{ij} \sigma_{(i+1)j}
- 
2J_1^\mathrm{L}  
\sum_i 
\sum_{\langle jk \rangle}^\mathrm{n.n.} 
\sigma_{ij} \sigma_{ik}
\nonumber \\
&-&
2J_2^\mathrm{L} 
\sum_i 
\sum_{ \langle jk \rangle}^\mathrm{n.n.n.} 
 \sigma_{ij} \sigma_{ik},
\\
\mathcal{H}_\mathrm{S}=
&-&2J_{\mathrm c}^{\rm S}\sum_{i,j} S_{ij}S_{(i+1)j}
\nonumber \\
&-&2J_{\mathrm 1}^{\rm S}\sum_i\sum_{\langle jk\rangle}^{\rm n.n.}
(1-\Delta (\sigma_{ij}-\sigma_{ik})^2) S_{ij}S_{ik}
\nonumber \\
&-&2J_{\mathrm 2}^{\rm S}\sum_i\sum_{\langle jk\rangle}^{\rm n.n.n.}
(1-\Delta (\sigma_{ij}-\sigma_{ik})^2) S_{ij}S_{ik}.
\end  {eqnarray}
The lattice part comes from the elastic energy: $(\sigma_{ij}-\sigma_{i'j'})^2$.
The spring constant is denoted by $J^\mathrm{L}_{(c,1,2)}$, where
each of ($c$, 1, 2) denotes a direction of the interaction:
$c$ denoting along the $c$-axis,
1 denoting the nearest-neighbor(n.n) pairs on the $c$-plane, and
2 denoting the next-nearest-neighbor(n.n.n) one on the $c$-plane.
The sign of the spring constant is determined based on the effect of the
exclusion volume effect.
It is positive along the $c$-axis: $J_c^{\rm L}>0$.
An ion pushes the next ion in the same direction.
The spring constant for the nearest pairs in the $c$-plane should be
negative: $J_1^{\rm L}<0$.
An ion shifts upward if the neighboring ion shifts downward, 
because ions try to stay away from the neighboring ions.
Therefore, there is frustration in the triangular lattice.
We choose $J^\mathrm{L}_2$ to be positive in order to realize
the $\uparrow$-$\uparrow$-$\downarrow$ state
observed experimentally at low temperatures.

The main idea of this paper is to make the spin-spin exchange integrals 
depend on the lattice variables.
We assume that
the interaction becomes weak if the exchange path is distorted.
Thus, the in-plane exchange interaction becomes $(1-\Delta)$ times weaker 
if $\sigma_{ij}$ and $\sigma_{ik}$ have opposite signs.
We use the same value of $\Delta$ for $J_1^{\rm S}$ and $J_2^{\rm S}$ for
simplicity.
The exchange path along the $c$-axis is rigid against the 
ion shift and hence we assume $J_c^{\rm S}$ to be unaffected.

We assume that the nearest-neighbor spin-spin interaction is 
antiferromagnetic ($J_1^{\rm S}<0$)
and the next-nearest-neighbor interaction
is ferromagnetic ($J_2^{\rm S}>0$)
in order to realize the ferrimagnetic state in the ground state.
The interactions along the $c$-axis in the real compound are 
antiferromagnetic.($J_c^{\rm S}<0$)

The lattice part and the spin part of the Hamiltonian have the same form of
the antiferromagnetic Ising model on the stacked triangular lattice.
They are connected by the $\Delta$ term of the form 
$-4J_{1,2}^{\mathrm S}\Delta \sigma_{ij}\sigma_{ik} S_{ij}S_{ik}$.
Thus,
the present model can be regarded as the Ashkin-Teller model.\cite{ashkin}

Experimental estimates of the exchange integrals were
$J_c^\mathrm{S}\simeq -62$ K, 
$J_1^\mathrm{S}\simeq -2.5$ K, and 
$J_2^\mathrm{S}\simeq 1$ K.\cite{morishitaC}
The estimate of $J_c^\mathrm{S}$ was obtained from the position of the broad 
maximum peak of $\chi_{\parallel}$.
We consider it underestimated,
which we  will discuss in Sec. \ref{sec:results}.

\subsection{Relaxation of frustration by lattice distortion}

\begin{figure}
\begin{center}
\includegraphics[width=5.0cm]{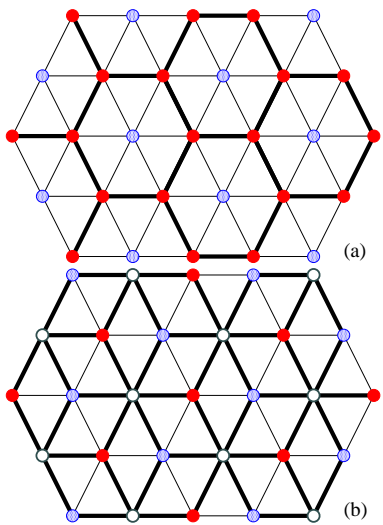}
\includegraphics[width=5.0cm]{fig2b.eps}
\end{center}
\caption{
(Color online)
Relaxation of frustration by the lattice distortion.
Red (solid) circles depict $c$-chains shifting upward.
Blue (grey) circles depict $c$-chains shifting downward.
Open (white) circles depict $c$-chains not shifting.
Thin (thick) lines depict weak (strong) interactions.
(a) When the lattice is deformed to a lattice-Ferri 
($\uparrow$-$\uparrow$-$\downarrow$) pattern, 
the spin-PD state is favored.
(b) When the lattice is deformed to a lattice-PD 
($\uparrow$-$\downarrow$-0) pattern, 
the spin-Ferri state is favored.
}
\label{fig:frustrate}
\end{figure}

Here, we consider how the ordered magnetic state is favored 
by the lattice distortion in the present spin-lattice model.
When the lattice takes the lattice-Ferri 
($\uparrow$-$\uparrow$-$\downarrow$) configuration,
the PD state of the spin system is favored magnetically.
As shown in Fig.~\ref{fig:frustrate}(a), the nearest-neighbor interactions 
between two $\uparrow$-shifted sublattices remain strong (depicted in 
the figure by thick lines), while
those between an $\uparrow$-shifted sublattice 
and a $\downarrow$-shifted sublattice are weakened by the $\Delta$ term.
The strong bonds form a honeycomb lattice, where
the spins are ordered antiferromagnetically.
The remaining spins on the $\downarrow$-shifted sublattice interact with
the spins on the honeycomb lattice through weak bonds.
The molecular field on the $\downarrow$-shifted sublattice from the
$\uparrow$-shifted sublattices
vanishes because of the antiferromagnetic ordering on the
$\uparrow$-shifted sublattices.
Then, the spins on the $\downarrow$-shifted sublattice may be disordered.
We call the PD state of the spin system as ``spin-PD" in this paper.

The similar argument is possible when the lattice system takes the lattice-PD 
($\uparrow$-$\downarrow$-0) state as shown in Fig.~\ref{fig:frustrate}(b).
The ferrimagnetic spin state is favored in this case.
The nearest-neighbor interactions between an
$\uparrow$($\downarrow$)-shifted sublattice 
and a 0-shifted sublattice are weakened by $\Delta/4$, while
those between the $\uparrow$-shifted sublattice and 
the $\downarrow$-shifted sublattice are weakened by $\Delta$.
The antiferromagnetic ordering is realized on the stronger bonds, which
is the ferrimagnetic state.
We call it in this paper ``spin-Ferri".
The present mechanism was discussed 
by Plumer {\it et al.},\cite{kawamura-spin-lattice}
when the lattice distortion is static.

\section{Monte Carlo Method}
\label{sec:method}

\begin{figure}
\begin{center}
\includegraphics[width=5.0cm]{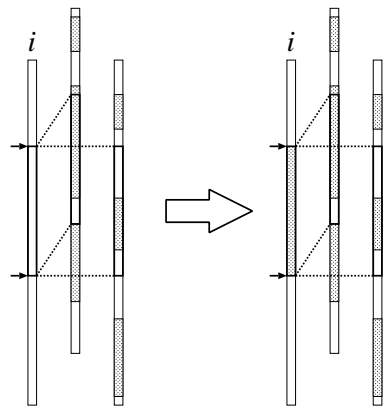}
\end{center}
\caption{
A schematic diagram of the MC updating procedure.
A cluster in an $i$-chain is defined between two edge arrows.
It is flipped using the sum of molecular fields from other chains.
}
\label{fig:method}
\end{figure}
\subsection{
Axial-bond-cluster flip algorithm
}

We briefly explain
our new Monte Carlo simulation algorithm.
The detail will be reported elsewhere.\cite{axcMC}

The origin of the slow MC dynamics in the quasi-one-dimensional 
($|J_c| \gg |J_1|$)
Ising system is a very long correlation length along the $c$-axis.
It rapidly grows at low temperatures as $\xi_c\sim \exp[|J_c|/T]$.
A large magnetic domain along the $c$-axis is not flipped by 
the standard single-spin-flip algorithm.
Koseki and Matsubara\cite{chb} 
introduced a cluster-heat-bath algorithm in order to solve this problem,
but it costs a long CPU time.
The possible size of simulations is restricted to
$|J_c/J_1|=10$, $N=36\times 36\times 360$, 
and $2\times 10^6$ MC steps.\cite{meloche}   

Here, we solve this problem using the
loop algorithm of quantum Monte Carlo (QMC) simulations.\cite{totaqmc}
In the QMC simulation, a $d$-dimensional quantum spin system is mapped
to a $(d+1)$-dimensional classical spin system\cite{suzukitrotter}
before actual simulations.
The additional dimension is called the Trotter direction.
Then, the classical spin system for the QMC algorithm can be interpreted 
as a stacked Ising model.
The Trotter direction of the QMC system is now the $c$-axis of the
stacked Ising model, and
the real-space directions of the QMC system are the $c$-plane of the
stacked Ising model.
The loop algorithm of the QMC simulation\cite{totaqmc} is to flip a ``loop", 
or an axial aligned-spin cluster along the Trotter direction in the QMC system.
Therefore, the algorithm can be readily applied to flip a correlated spin
cluster along the $c$-axis of the stacked Ising system.

The size of the cluster is a stochastic variable in each update of the
cluster algorithm.
Using a proper probability we generate locations of the cluster edges,
which are memorized in the computer array.
Then, we calculate the sum of molecular fields from other spins to the cluster 
between two neighboring edges (two arrows in Fig.~\ref{fig:method}).
The cluster is flipped using the heat-bath probability by this molecular field.

We noticed that the
required computational procedures and the amount of computer memory 
are independent of the correlation length $\xi_c$.
We do not need to memorize spin states of all sites.
Locations of the cluster edges and the spin state at each chain edge are 
stored and utilized in the simulation.
The linear size along the $c$-axis, $L_c$, is set to
$\xi_c$ times larger than $L$.
The system with $L^2 \times \xi_c L$ spins is simulated with
an effort of $L^3$.
The new algorithm becomes exponentially efficient at low temperatures.
In this paper, we set $L=104$ for all data.  
An effective spin number at low temperatures exceeds $10^8$.
The periodic boundary conditions are imposed on the lattice.

\subsection{
Observables
}
We observe in the present MC simulations the following physical quantities:
the sublattice order parameters, 1/3-structure factors, 1-structure factors,
and the uniform magnetic susceptibility.
The sublattice order parameters are the sublattice polarization,
$m_{\eta}^\mathrm{L}$,
and the sublattice magnetization,$m_{\eta}^\mathrm{S}$, respectively.
They are defined as
\begin{eqnarray}
m_{\eta}^\mathrm{L}&=&\frac{1}{N_\mathrm{sub}}\sum_i\sum_{j\in \eta} 
\sigma_{ij}
\\
m_{\eta}^\mathrm{S}&=&\frac{1}{N_\mathrm{sub}}\sum_i(-1)^i\sum_{j\in \eta} 
S_{ij},
\end  {eqnarray}
where $\eta= \alpha, \beta, \gamma$ denotes one of three sublattices in the
triangular lattice,
and $N_\mathrm{sub}\equiv N/3$.

The following structure factors are defined 
in order to detect phase transitions and to compare with
the neutron experimental data:
\begin{eqnarray}
(f_{1/3}^\mathrm{L})^2&=&\frac{1}{{8}}
\left\langle
\sum_{\eta=\alpha, \beta, \gamma} 
(m_{\eta}^\mathrm{L} - m_{\eta+1}^\mathrm{L})^2
\right\rangle,
\\
(f_{1/3}^\mathrm{S})^2&=&\frac{1}{{8}}
\left\langle
\sum_{\eta=\alpha, \beta, \gamma} 
(m_{\eta}^\mathrm{S} - m_{\eta+1}^\mathrm{S})^2
\right\rangle,
\label{eq:f3s}
\\
(f_{1}^\mathrm{L})^2&=&\left\langle
(m_{\alpha}^\mathrm{L}+m_{\beta}^\mathrm{L}+m_{\gamma}^\mathrm{L})^2
\right\rangle,
\\
(f_{1}^\mathrm{S})^2&=&\left\langle
(m_{\alpha}^\mathrm{S}+m_{\beta}^\mathrm{S}+m_{\gamma}^\mathrm{S})^2
\right\rangle.
\label{eq:f1s}
\end  {eqnarray}
The 1/3-structure factor takes a finite value when the ferrimagnetic state
or the PD state is realized. 
It detects the phase transition between the PD phase and the paramagnetic phase.
The phase transition between the PD phase and the ferrimagnetic phase 
is detected by $f_1$.

\subsection{Mean-field-like treatment of MC update}

In the present simulation, 
spin variables and lattice variables are updated separately and alternatively.
In the calculation of the heat-bath update probability, we use the
following approximation to simplify the simulation.
For an update of a spin variable $S_{ij}$, 
we calculated the four-body energy,
$
-4J_{1,2}^{\rm S}\Delta \sigma_{ij}\sigma_{ik}S_{ij}S_{ik}
$
by replacing the lattice variable $\sigma_{ij}$ with a mean value
$\bar{\sigma_j}\equiv \sum_{i=1}^{Lc}\sigma_{ij}/L_c$,
namely as
$
-4J_{1,2}^{\rm S}\Delta (\bar{\sigma_{j}}\bar{\sigma_{k}})S_{ij}S_{ik}.
$
For an update of a lattice variable $\sigma_{ij}$, 
we replaced the spin variable with a mean value 
$\bar{S_j}\equiv \sum_{i=1}^{Lc}S_{ij}/L_c$,
namely as
$
-4J_{1,2}^{\rm S}\Delta (\bar{S_{j}}\bar{S_{k}}) \sigma_{ij}\sigma_{ik}.
$

This mean-filed treatment may be justified by the following argument.
In a cluster updating procedure,
the sum of the molecular field to a cluster
is calculated to estimate the updating probability.
Since the cluster size, $\sim \exp[|J_c^{\rm S,L}|/T]$, 
is very large around/below the critical temperature,
the mean over a cluster can be approximated by the mean over the whole chain.

The above mean-field treatment possibly influences the critical properties
of the phase transitions.
Since our main purpose here is to explain the experimental results,
most of the simulations are carried out in the off-critical regions.
Therefore, we consider that this mean-field treatment does not affect 
our numerical results in the present paper.
The investigations on the critical properties are left for future study.

\subsection{Simulation conditions}

The choice of the initial state is important in the present simulation.
Since each of the lattice system and the spin system exhibit 
two successive transitions,
we have several possible combinations of ordering patterns as we change
the temperature.
We therefore used the mixed phase 
initialization,\cite{mix-state-start1,mix-state-start2,mix-state-start3}
where
we prepare several initial spin-lattice states and spatially mix them.
For example, we start the simulation with the following initial state
when the temperature is near the spin-PD transition temperature.
The lattice system above this temperature takes the
lattice-PD ($\uparrow$-$\downarrow$-0) state.
Because the spin-PD state favors the lattice-Ferri
($\uparrow$-$\uparrow$-$\downarrow$) state (Fig.~\ref{fig:frustrate}(a)),
they may appear at the same temperature.
Therefore, a half of the system is set to the spin-PD state and the
lattice-Ferri state,
while the other half is set to the
spin-paramagnetic state and the lattice-PD state.
The former one appears below the transition temperature, while the
latter appears above it.
We tried other choices of mixed states and verified the equilibration.

The typical number of initialization MC steps was one thousand and
that of total MC steps was ten thousands.
It is sufficient except for the vicinity of the transition temperature.
We performed thirty independent MC runs and took the average over these runs.

\section{Results}
\label{sec:results}

\subsection{Requirements from the experiments}

\begin{figure}
\begin{center}
\includegraphics[width=8.0cm]{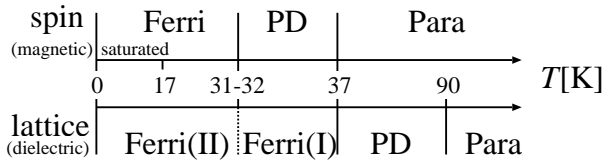}
\end{center}
\caption{
A phase diagram of RbCoBr$_3$.
Solid lines show the known phase transition temperatures.
A broken line at 32 K is a new structural transition found in this paper.
}
\label{fig:phase}
\end  {figure}

Experimental findings are listed in the following.
They should be reproduced by the simulations.
\begin{enumerate}
\item[(i)]
The uniform magnetic susceptibility shows a broad peak at $T=100$ K.
\cite{experi-mag}
\item[(ii)]
The dielectric constant shows a small anomaly at $T=90$ K,
\cite{morishita52,experi-mag}
where the lattice-PD ($\uparrow$-$\downarrow$-0)
state is considered to appear.
\item[(iii)]
As the temperature decreases from the room temperature,
the first magnetic phase transition occurs at $T_{\rm N1}=37$ K.
The neutron-scattering data of (1/3 1/3 1), which corresponds to 
$(f_{1/3}^{\rm S})^2$ of Eq.~(\ref{eq:f3s}), show a rapid increase 
below this temperature, while those of (1 1 1),
which corresponds to $(f_{1}^{\rm S})^2$ of Eq.~(\ref{eq:f1s}), remains zero.
The spin-PD state is considered to appear at this temperature.
\item[(iv)]
The dielectric constant increases below $T_{\rm N1}=37$ K.
The lattice-Ferri ($\uparrow$-$\uparrow$-$\downarrow$) state is 
considered to appear.\cite{morishita52,experi-mag}
\item[(v)]
The neutron-scattering data of (1 1 1) begin to increase at $T_{\rm N2}=31$ K.
It is considered as the second magnetic phase transition as the
temperature decreases from the room temperature.
The temperature dependence of the (1 1 1) 
data is linear with $T$.\cite{experi-f}
\item[(vi)]
All the neutron data saturate at $T=17$ K, below which the magnetic 
order seems to be perfect.\cite{experi-f}
\item[(vii)]
The dielectric constant also shows an anomaly at 32 K.\cite{experi-mag}
The temperature is very close to $T_{\rm N2}$.
It is not known whether it is another structural phase transition or not.
\end  {enumerate}
The above experimental evidences are summarized in Fig.~\ref{fig:phase}.

The requirement (i) determines the energy scale of $J_c^{\rm S}$,
the requirement (vi) determines $J_2^{\rm S}$,
the requirement (iii) ($T_{\rm N1}$) determines the ratio 
$J_2^{\rm S}/J_1^{\rm S}$,\cite{shiba,axcMC}
and
the requirement (ii) determines the ratio $J_2^{\rm L}/J_1^{\rm L}$.
The other parameters are
determined by the temperature dependence
of the structure factor between 20 K and 37 K.

\subsection{The spin-lattice model}

\begin{figure}
\begin{center}
\includegraphics[width=7.0cm]{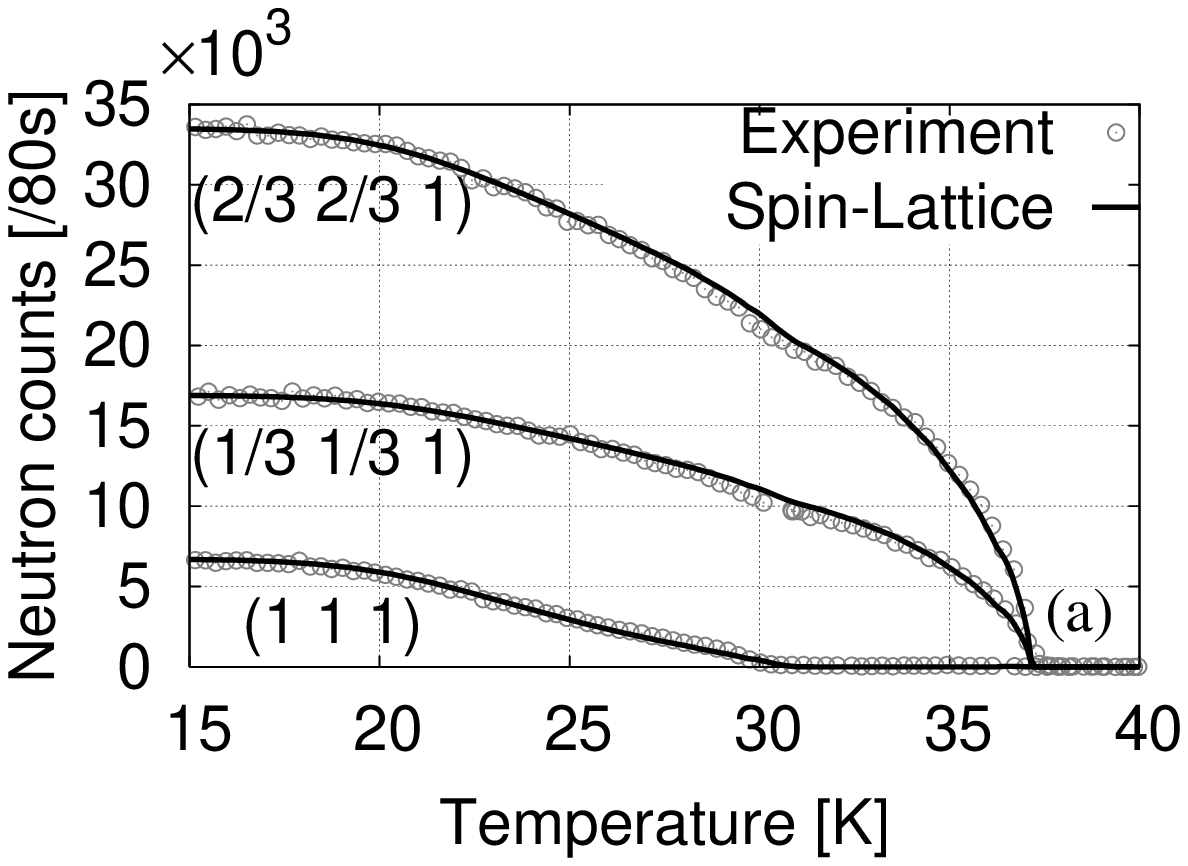}
\includegraphics[width=7.0cm]{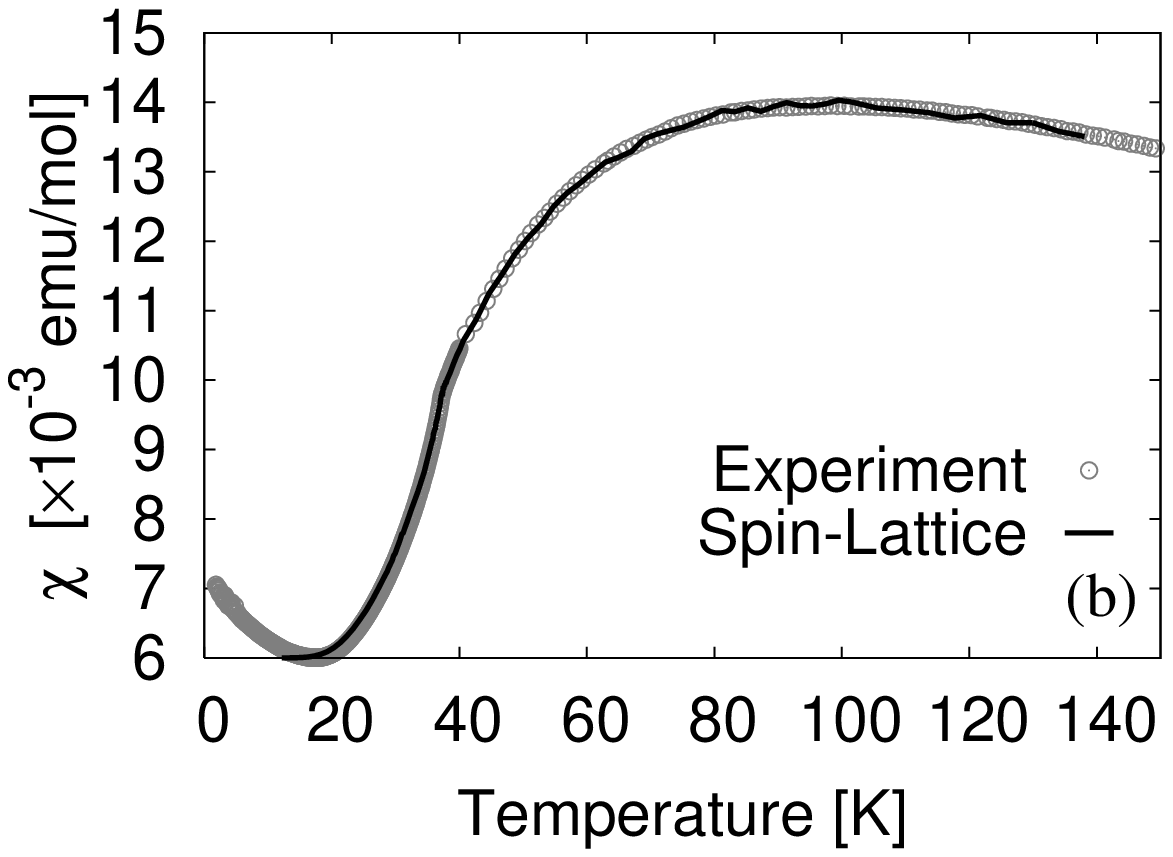}
\end{center}
\caption{
Results of the spin-lattice model.
The lattice parameters are set to the values in Eq.~(\ref{eq:73K}).
(a) The MC data of the structure factor are compared with the
neutron experimental data.\cite{experi-f}
The MC data of $(f_1^{\rm S})^2$ are multiplied to
coincide with the experimental data of (1 1 1).
Those of $(f_{1/3}^{\rm S})^2$ are multiplied to coincide with the 
experimental data of (1/3 1/3 1) and (2/3 2/3 1).
(b) The MC data of the uniform magnetic susceptibility are compared with
the experimental data.\cite{experi-mag}
The amplitudes of the simulation data
and the constant contribution from the non-magnetic impurity are
determined so that the maximum value and the minimum 
value agree with the experimental data.
}
\label{fig:spin-lattice}
\end{figure}

\begin{figure}
\begin{center}
\includegraphics[width=7.0cm]{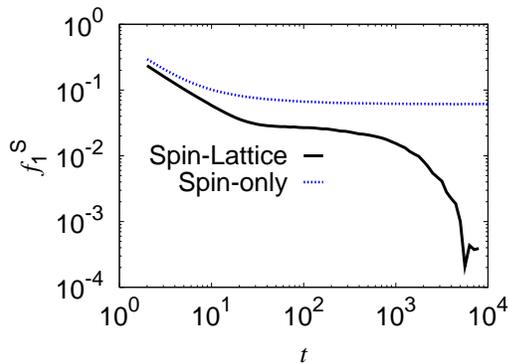}
\end{center}
\caption{
A nonequilibrium relaxation plot of the magnetic 
structure factor $f_1^{\rm S}$ when
the simulation starts from the ferrimagnetically-ordered state.
The temperature is 35.5 K, just below $T_{\rm N1}=37$~K.
The relaxation function of the spin-only system converges to a finite value,
while that of the spin-lattice system decays exponentially.
The lattice parameters are those of Eq.~(\ref{eq:73K}).
}
\label{fig:ner}
\end{figure}

Numerical results are shown in Fig.~\ref{fig:spin-lattice}.
The seven parameters are determined
in order to fit the neutron data by visual inspection.
The spin parameters are uniquely determined as
\begin{equation}
J_c^{\rm S}=-97~{\rm K},~
J_1^{\rm S}=-2.4~{\rm K},~
J_2^{\rm S}= 0.14~{\rm K}.
\label{eq:spinJ}
\end  {equation}
Those for the lattice system were not uniquely determined.
There are several choices that reproduce the experimental results.
We present two choices of the lattice parameters in this paper.
Other possible parameter choices range between these two estimates.
They are
\begin{equation}
J_c^{\rm L}= 73~{\rm K},~
J_1^{\rm L}=-49~{\rm K},~
J_2^{\rm L}= 0.38~{\rm K},~
\Delta=0.20,
\label{eq:73K}
\end  {equation}
and
\begin{equation}
J_c^{\rm L}= 61~{\rm K},~
J_1^{\rm L}=-57~{\rm K},~
J_2^{\rm L}= 0.61~{\rm K},~
\Delta=0.24.
\label{eq:61K}
\end  {equation}
Both parameter choices reproduce the neutron data and the 
susceptibility data fine.
Quality of the fitting is the same within the visual inspection.
These parameters reproduce
the susceptibility data from 20 K to 140 K, 
including a convex change at 37 K.
The intermediate phase between 31 K and 37 K is identified as the spin-PD phase,
because
the nonequilibrium relaxation data in Fig.~\ref{fig:ner} exhibits that 
$f_1^{\rm S}$ disappears exponentially.

We cannot uniquely determine the lattice parameters because of 
lack of information that determines $J_c^{\rm L}$.
In the present lattice system,
there is not an observable corresponding to the magnetic susceptibility,
which determines $J_c^{\rm S}$.
For each choice of $J_c^{\rm L}$, we can find $J_1^{\rm L}$, $J_2^{\rm L}$,
and $\Delta$ in order to satisfy the experimental data.
The ratio $J_c^{\rm S}/J_c^{\rm L}$ takes a value from 1.6 to 1.0, 
which agrees with the previous paper.\cite{shiramaster}

\begin{figure}[t]
\begin{center}
\includegraphics[width=8.0cm]{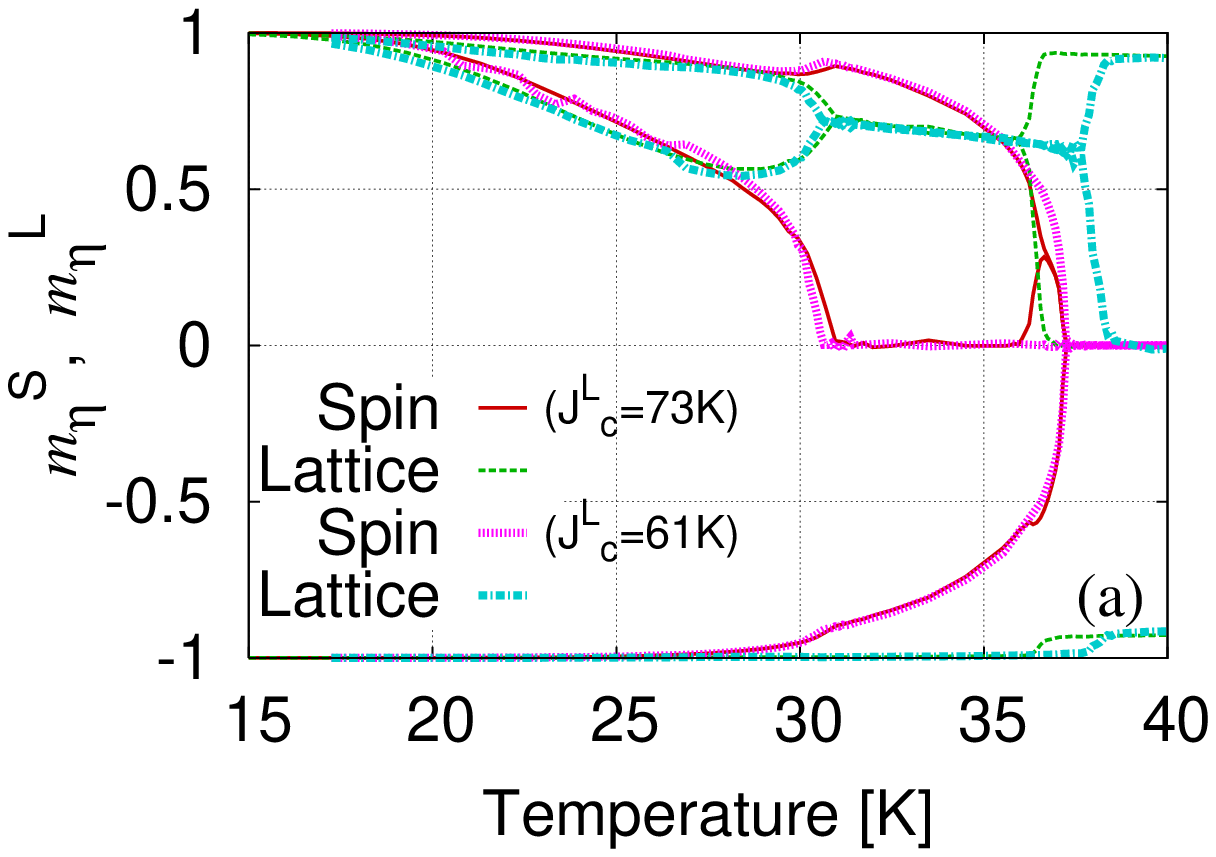}
\includegraphics[width=8.0cm]{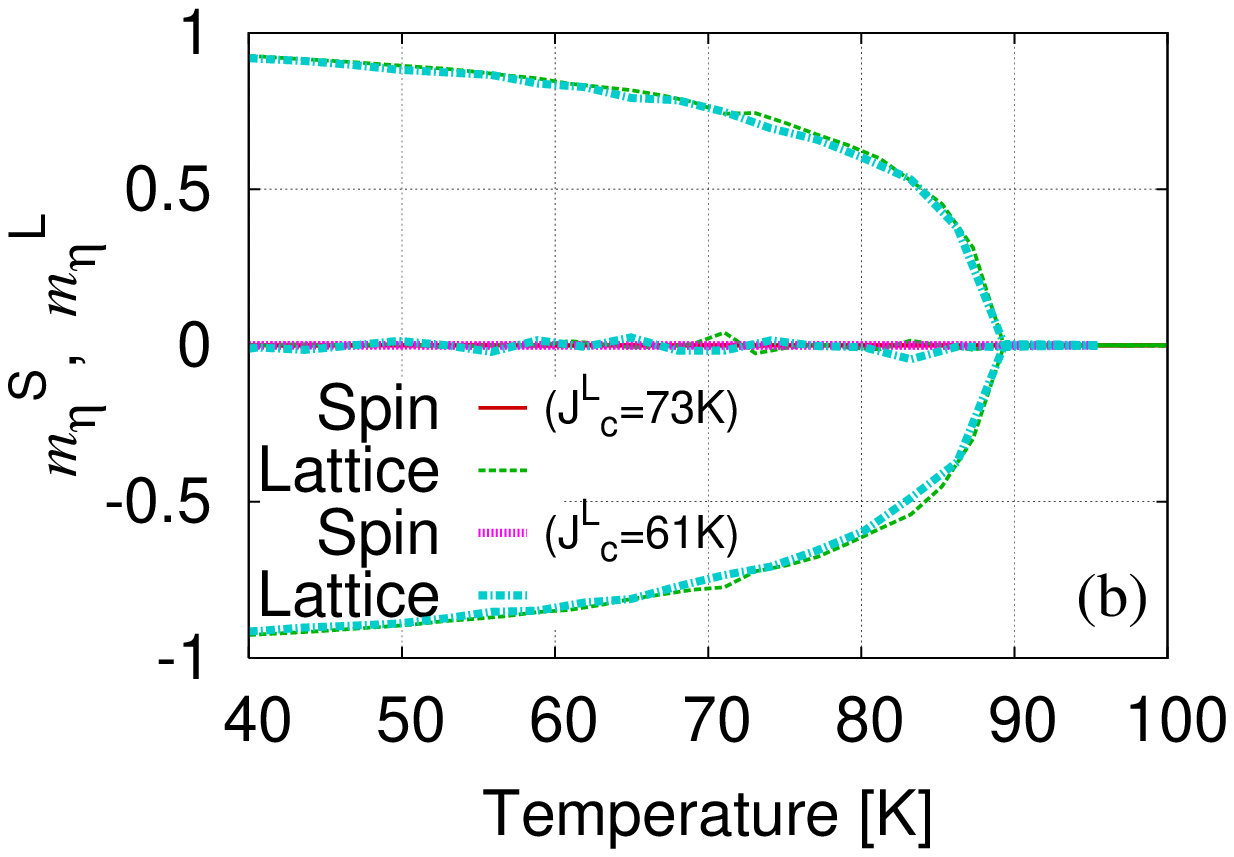}
\end{center}
\caption{
(Color online)
Sublattice profiles of spin variables and lattice variables.
The amplitude is normalized to unity when the sublattice order is perfect.
Thin (red and green) lines are results of the $J_c^{\rm L}=73$ K parameter set,
while
thick (magenta and light-blue) lines are results of the $J_c^{\rm L}=61$ K 
parameter set, respectively.
(a) The temperature ranges from 15 K to 40 K, where
the corresponding neutron experiments were performed.
(b) The high-temperature data.
}
\label{fig:sub}
\end{figure}

We have observed the sublattice magnetization and the
sublattice polarization.
Figures \ref{fig:sub}(a) and \ref{fig:sub}(b) 
show the temperature dependence of the profiles.
As shown in Fig.~\ref{fig:sub}(b),
the structural phase transition occurs at $T\simeq 90$ K, 
below which the lattice system
takes the lattice-PD ($\uparrow$-$\downarrow$-0) state.
It reproduces the experimental requirement (ii).
The spin system remains paramagnetic in this temperature region.

The spin transition and the lattice transition occur near 
37 K (Fig.~\ref{fig:sub}(a)).
They are not always simultaneous.
When $J_c^{\rm L}$ is set to 73 K,
the spin transition occurs at 37.2 K, which is slightly higher than 
the lattice transition temperature, 36.8 K.
On the other hand, the lattice transition temperature becomes 
38.5 K when $J_c^{\rm L}$ is set to 61 K, while the spin transition 
temperature remains the same.
The experimental finding of the simultaneous magneto-dielectric phase
transition at 37 K is an accidental coincidence.
The two transition may occur at slightly different temperatures.

In the case of $J_c^{\rm L}= 73 $K, 
a weak ferrimagnetic state appears at 37.2 K.
This is because the lattice takes the lattice-PD 
($\uparrow$-$\downarrow$-0) state at this temperature, and it
favors the spin-Ferri state (Fig.~\ref{fig:frustrate}(b)).
When the lattice transition occurs at 36.8 K,
the spin-Ferri state disappears and the spin-PD state appears
because the lattice-Ferri state favors the spin-PD state
(Fig.~\ref{fig:frustrate}(a)).
This is an outcome of the spin-lattice coupling.
The situation changes when $J_c^{\rm L}$ is set to 61 K.
The lattice transition occurs at 38.5 K. 
Since the lattice system takes the lattice-Ferri
($\uparrow$-$\uparrow$-$\downarrow$) state at 37.2 K,
the favored spin order is the spin-PD.
The weak spin-Ferri state does not appear in this case.
This weak ferrimagnetic phase is so narrow that it may not
be observed in experiments if it exists.

It is noticed that the spin transition temperature is robust against
the change of the lattice parameters.
As far as we observed in the MC simulations with several lattice 
parameter choices, the spin transition temperature 
from the paramagnetic phase to the intermediate phase
always occurs at 37 K.
The spin transition temperature seems to be determined by the spin parameters
Eq.~(\ref{eq:spinJ}).
On the other hand, the lattice system controls the type of the spin order 
at this temperature.

In the intermediate phase,
the amplitudes of two $\uparrow$-shifted sublattice polarization and that of
one $\downarrow$-shifted sublattice polarization are different.
The former one is not saturated, while the latter is saturated.
It is the two-sublattice lattice-Ferri 
($\uparrow$-$\uparrow$-$\downarrow$) state (Fig.~\ref{fig:lattice}(b)).
An increase of the $\uparrow$-shifted polarization is slow and almost
linear with the decreasing temperature.

The low-temperature magnetic transition occurs at 31 K, below which
the ferrimagnetic state appears.
Each sublattice magnetization takes a different value.
An inversion symmetry between a spin-up sublattice and a spin-down sublattice
is broken.
It is the three-sublattice ferrimagnetic state, which was predicted
to appear by the mean-field approximation.\cite{nishi-todo.MF}
As the temperature decreases, the sublattice magnetizations 
approach the unity and the 
perfect ferrimagnetic order is realized at temperatures near $T=17$ K.
This saturation temperature depends on $J_2^{\rm S}$.

The lattice transition always occur 
at the low-temperature magnetic transition temperature.
It is the simultaneous spin-lattice phase transition.
Above this spin-lattice transition temperature,
the lattice system takes the two-sublattice lattice-Ferri state
(Fig.~\ref{fig:lattice}(b)), which is the ground-state configuration.
When the spin-Ferri state appears below the transition temperature, 
the lattice state is
deformed toward the lattice-PD ($\uparrow$-$\downarrow$-0) state because the
spin-Ferri state favors the lattice-PD state.
A combination of the partial lattice-Ferri state and the partial
lattice-PD state yields the three-sublattice lattice-Ferri state
(Fig.~\ref{fig:lattice}(d)).
The space group changes from $P\bar{3}c1$ to $P3c1$.
It is a clear evidence for a strong correlation between the spin system
and the lattice system.
The lattice system alone cannot make this structural phase transition,
because the lattice system is in the ground-state phase above the
transition temperature.
Therefore, this may be the spin-lattice transition 
driven by the spin degrees of freedom.

\begin{figure}
\begin{center}
\includegraphics[width=7.0cm]{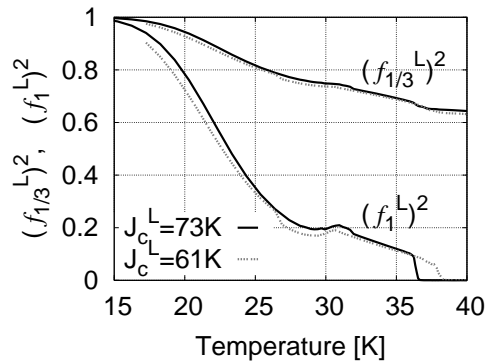}
\end{center}
\caption{
The structure factor data of the lattice system.
The data of $(f_1^L)^2$ are multiplied by 9.
}
\label{fig:f-lattice}
\end{figure}

Figure \ref{fig:f-lattice} shows the MC results of the
structure factor of the lattice system.
The $f_{1/3}$-structure factor shows small anomalies at 37 K and 31 K.
The temperature dependence between 37 K and 31 K is slow and 
almost linear with $T$.
The $f_1$-structure factor data qualitatively agree with the
experimental results of the spontaneous polarization.\cite{experi-mag}
The data show a decrease below 31 K because the lattice-Ferri state
is deformed toward the lattice-PD state by the ferrimagnetic transition.
In the real experiment,\cite{experi-mag} 
the spontaneous polarization shows a minimum at 
23 K, while the minimum occurs at 28 K in the present simulation.

\subsection{The spin-only model}

\begin{figure}
\begin{center}
\includegraphics[width=7.0cm]{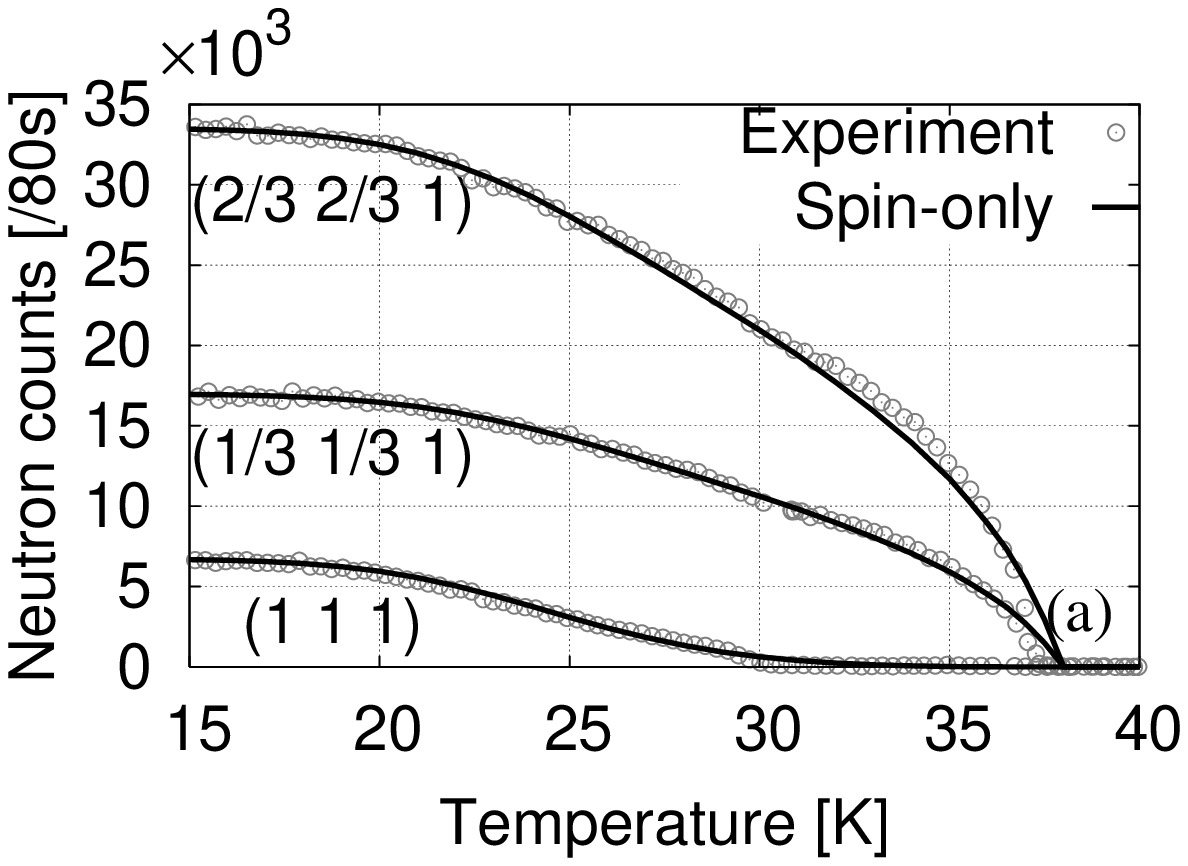}
\includegraphics[width=7.0cm]{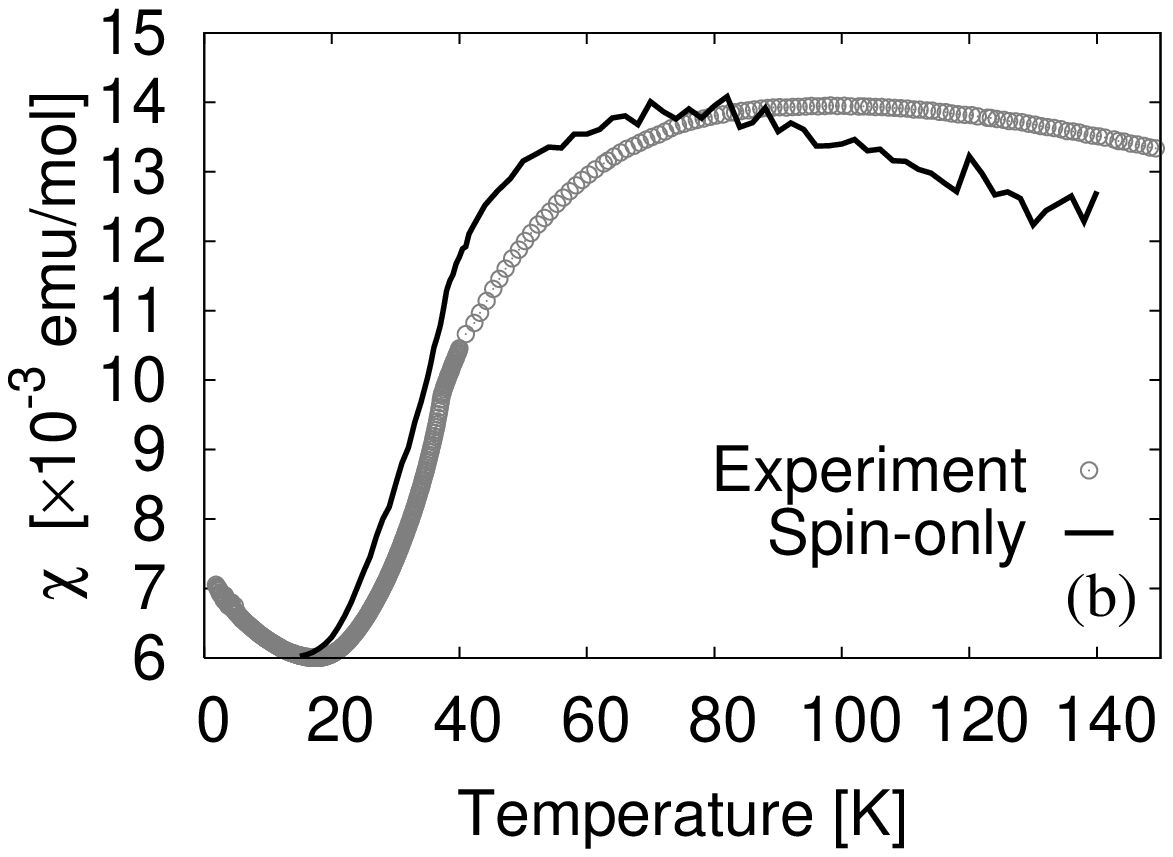}
\end{center}
\caption{
Results of the spin-only model.
(a) The MC data of the structure factor are compared with the
neutron experimental data.\cite{experi-f}
The MC data of $(f_1^{\rm S})^2$ are multiplied to
coincide with the experimental data of (1 1 1).
Those of $(f_{1/3}^{\rm S})^2$ are multiplied to coincide with the 
experimental data of (1/3 1/3 1) and (2/3 2/3 1).
(b) The MC data of the uniform magnetic susceptibility are compared with
the experimental data.\cite{experi-mag}
The amplitudes of the simulation data
and the constant contribution from the non-magnetic impurity are
determined so that the maximum value and the minimum 
value agree with the experimental data.
}
\label{fig:spinonly}
\end{figure}

We show that the spin-only system cannot explain all the experimental data.
The three parameters, 
$J_c^{\rm S}$,
$J_1^{\rm S}$,
and $J_2^{\rm S}$,
are determined in order to fit the neutron experimental data fine
by visual inspection.
Figure \ref{fig:spinonly} shows the result.
The estimates are
\[
J_c^{\rm S}=-77~{\rm K},~
J_1^{\rm S}=-3.8~{\rm K},~
J_2^{\rm S}= 0.58~{\rm K}.
\]
Agreement with the neutron data is good, while
the susceptibility data disagree with the experiment significantly.
If we choose estimates that fit the susceptibility data,
the neutron data disagree in turn.
We cannot find estimates that satisfy both experimental data at the same time.

The MC data of the structure factor of (1 1 1) take very small but 
finite values between 31 K and 37 K, as shown in Fig.~\ref{fig:ner}.
It suggests that the ferrimagnetic order is finite and
the intermediate spin-PD phase disappears.
The direct transition from the paramagnetic phase to the ferrimagnetic
phase is an outcome of a rather large estimate of $J_2^{\rm S}$.

The spin-only model does not consider the dielectric characteristics of 
RbCoBr$_3$.
We cannot explain the successive structural phase transitions by 
this model.
Using three parameters,
$J_c^{\rm S}$, $J_1^{\rm S}$, and  $J_2^{\rm S}$, we only reproduce 
either the neutron experimental data or the magnetic susceptibility data.
On the other hand, the spin-lattice model explains quantitatively 
both dielectric and magnetic properties using seven parameters.
It suggests that
the interplay between the spin system and the lattice system is essential
in this compound.

\subsection{Perturbations}

We consider perturbation effects to the present model.
It is intended to see how robust the characteristic behaviors of RbCoBr$_3$
are against perturbations as well as to propose
further experimental investigations.
Here, we consider three perturbations.
The former two perturbations couple with the lattice system, while the
last one changes the spin-lattice coupling parameter.

\begin{figure}
\begin{center}
\includegraphics[width=8.0cm]{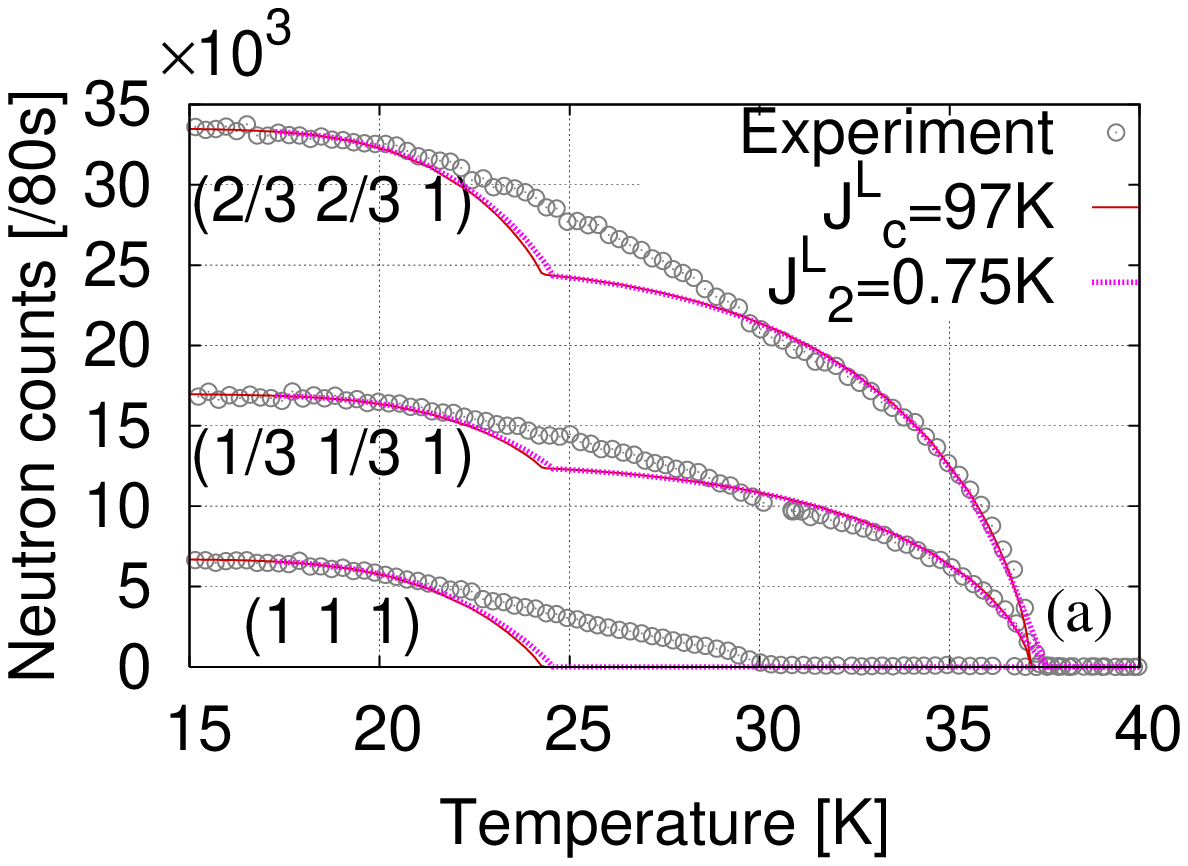}
\includegraphics[width=8.0cm]{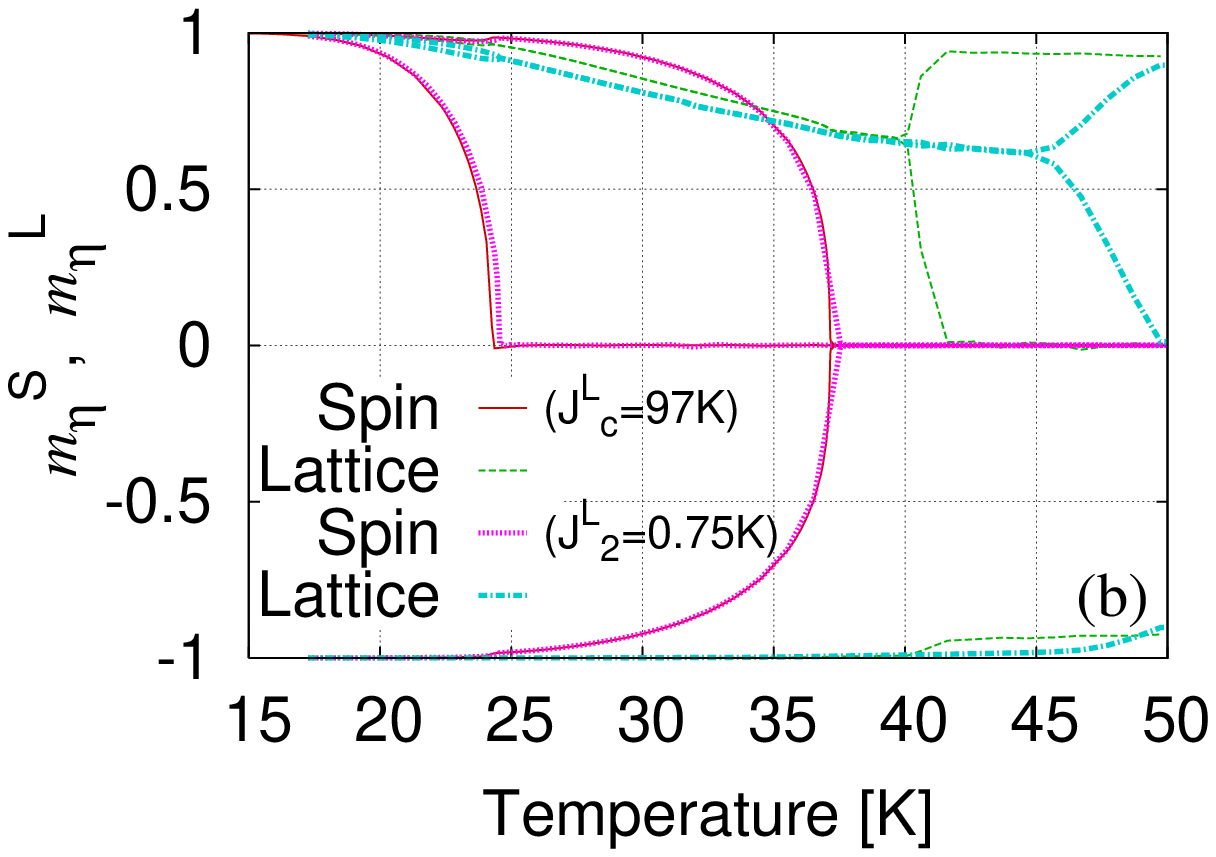}
\end{center}
\caption{
(Color online)
Perturbations on exchange interactions.
In one case, we increased $J_c^{\rm L}$ from 73 K to 97 K, 
(thin, red and green)
In the other case, we increased
 $J_2^{\rm L}$ from 0.38 K to 0.75 K (thick, magenta and light-blue).
The other parameters are the same as in Eq.~(\ref{eq:73K}).
(a) Neutron data of RbCoBr$_3$ are compared with the MC data
of the structure factor in the perturbed cases.
The MC data are multiplied so that the saturation value coincide
with the experimental data at 17~K.
(b) Sublattice profiles of the spin and the lattice variables.
}
\label{fig:largeJcL}
\end{figure}

First, a lattice interaction parameter is changed in order to make
the lattice system hard.
It corresponds to a pressure effect.
We increase the interaction (spring constant) along the $c$-axis, $J_c^{\rm L}$,
from 73 K to 97 K, while the other parameters remain the same 
as in Eq.~(\ref{eq:73K}).
Figure \ref{fig:largeJcL} shows the results.
The lattice transition to the 
two-sublattice $\uparrow$-$\uparrow$-$\downarrow$ state occurs at 40 K
(thin green lines in Fig.~\ref{fig:largeJcL}(b)),
while it occurs at 37.2 K (thin green line in Fig.~\ref{fig:sub}(a))
in the unperturbed case.
The magnetic transition occurs at 37.1~K 
(red and magenta lines in Fig.~\ref{fig:largeJcL}(b)), and the PD state appears.
This magnetic transition temperature is robust against the lattice 
perturbation.
The PD phase continues to the lower temperature, and the ferrimagnetic
transition occurs at 24 K.
We observe the small bifurcation of the lattice profiles at this temperature.
The two-sublattice lattice-Ferri state is deformed very weakly to the
three-sublattice lattice-Ferri state.
The simultaneous spin-lattice transition also occurs in the perturbed system.
The PD phase becomes wider as in the typical ABX$_3$ compounds.
Another clear difference from the original parameter set is that
the (1 1 1) structure factor is convex when it appears at 24~K,
while the original one is linear.

Second, we changed $J_2^{\rm L}$ from 0.38 K to 0.75 K, 
while the other parameters remain the same as in Eq.~(\ref{eq:73K}).
This perturbation favors the lattice-Ferri ($\uparrow$-$\uparrow$-$\downarrow$)
state.
It may correspond to applying the electric field to this compound.
The results of the spin system are the same as the case where we
changed $J_c^{\rm L}$.
As shown in the figures, both perturbations produce 
the same temperature dependences of the spin profiles.
On the other hand, the lattice profiles are different.
The lattice $\uparrow$-$\uparrow$-$\downarrow$ transition 
occurs at 45 K (thick light-blue lines in Fig.~\ref{fig:largeJcL}(b)), 
while it occurs at 40 K in the $J_c^{\rm L}$ perturbation.
Experiments under the electric field or the high pressure may detect 
these changes.

\begin{figure}
\begin{center}
\includegraphics[width=7.0cm]{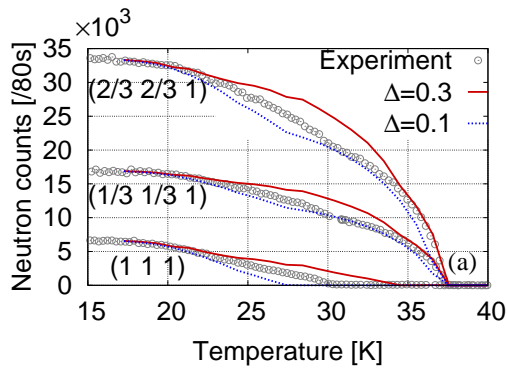}
\end{center}
\caption{
(Color online)
Neutron data of RbCoBr$_3$ are compared with the MC data of the
structure factor when $\Delta$ is set to 0.1 and 0.3. 
The other parameters are the same as in Eq.~(\ref{eq:73K}).
The MC data are multiplied so that the saturation value coincide
with the experimental data at 17~K.
}
\label{fig:DeltaLS}
\end{figure}

The last perturbation changes the spin-lattice coupling parameter.
We set $\Delta=0.1$ and $0.3$ while the other parameters are unchanged
from Eq.~(\ref{eq:73K}).
Figure \ref{fig:DeltaLS} shows the results of the structure factor
compared with the neutron experimental data.
The PD transition temperature and the saturation temperature are robust
against this perturbation.
On the other hand,
the ferrimagnetic transition temperature depends on $\Delta$,
which is observed by the change of slope of the (1 1 1) data.
As we increase $\Delta$, the ferrimagnetic transition temperature increases.
The spin-lattice coupling relaxes frustration and
stabilizes the ferrimagnetic state.

Comparing the above with the results of the original parameters 
(Figs.~\ref{fig:spin-lattice} and \ref{fig:sub})
we notice that there are several spin-lattice effects.
The ferrimagnetic state appears at higher temperatures because of 
the relaxation of frustration.
The (1 1 1) structure factor linearly depends on the temperature.
There appears a three-sublattice $\uparrow$-$\uparrow$-$\downarrow$ state.
These characteristic behaviors in RbCoBr$_3$
are fragile and disappear when the spin-lattice coupling changes.
Control of the lattice system by the electric field or the pressure 
may produce a new effect to the spin system.

\section{Discussion}
\label{sec:discussion}

The successive phase transitions of RbCoBr$_3$ are well explained by the 
spin-lattice model introduced in this paper.
Numerical data of our Monte Carlo simulations quantitatively 
agree with the experimental results.
The spin-lattice coupling is found to be essential in this system.
It produces nontrivial behaviors of RbCoBr$_3$ different from other typical
ABX$_3$ compounds.
The present analysis was enabled by the new cluster flip algorithm, 
which eliminates the slow MC dynamics in the quasi-one-dimensional 
frustrated spin system.

The magneto-dielectric transition at 37 K is not always simultaneous.
It is a coincidence that the spin transition and the lattice transition
occur at the close temperatures in RbCoBr$_3$.
They may differ if the interaction parameters are different.
The spin transition temperature is possibly determined independently from the
lattice system.
The spin-lattice coupling only determines what type of the spin order 
is realized below this transition temperature.

On the other hand, the magneto-dielectric transition at 31 K 
is always simultaneous.
It is the spin-driven lattice transition.
The lattice symmetry changes in order to realize the ferrimagnetic state.
Therefore, the anomaly of the spontaneous polarization observed experimentally
\cite{experi-mag} at 32 K is considered 
as an indication of another structural transition,
where the space group changes from $P\bar{3}c1$ to $P3c1$.
Further experiments to ensure this theoretical prediction is expecting.
An anomaly at 9 K observed experimentally has not been identified within the
present spin-lattice model.

The criticality of the phase transitions is an interesting future problem.
The linear temperature dependence of the (1 1 1) 
structure factor below 31 K in Fig.~\ref{fig:spin-lattice}(a) may be an
indication of the mean-field universality $\beta=1/2$.
Though the interaction range is limited to the second-nearest neighbor,
the spin-lattice coupling effectively makes it long-ranged because of
the large correlation lengths of both spin variables and lattice variables.
This mean-field criticality is supported by a model proposed
recently by Miyashita {\it et al.}\cite{miyashita-spincrossover}
Their model for a magnetic phase transition in spin-crossover materials
is similar to our spin-lattice model.
In their model,
the spin takes either a high-spin state or a low-spin state.
The volume of a magnetic ion depends on the spin state,
which produces an effective spin-lattice coupling.
They observed the mean-field universality by the detailed scaling analysis 
on the model system.
If the mean-field universality appears in RbCoBr$_3$,
it may be observed in other magneto-dielectric compounds, e.g., 
$R$Fe$_2$O$_4$.\cite{kakurai,ishihara}

It should be commented that our mean-field-like treatment of the 
MC updating may have affected the critical phenomenon.
This treatment averages the lattice variables along the chain.
It may produce effective long-range spin-spin correlations along the chain.

In the present model, the lattice parameters have not been determined
uniquely.
The lattice model is simple, taking only the elastic energy into account.
Our assumption of the spin-lattice coupling only models the 
deformation of the lattice system 
as an influence to the spin system.
Some modifications to the model may be necessary when we discuss the 
magnetic-dielectric cross correlation under an
electric field and a magnetic field.

\acknowledgments

The use of random number generator RNDTIK programmed by
Prof. N. Ito and Prof. Y. Kanada is gratefully acknowledged.
An author TN thanks Dr. Y. Konishi for fruitful discussions, and
thanks Prof. N. Hatano for the critical readings of the manuscript.


\begin{thebibliography}{99}

\bibitem{HFM2006}
For example,
{\it 
Proceedings of the International Conference on Highly Frustrated Magnetism, Osaka, Japan, 15-19 August 2006
},
J. Phys.: Condens. Matter \textbf{19} No 14 (11 April 2007) 

\bibitem{CsCoBr3}
W. B. Yelon, D. E. Cox, and M. Eibsch\"utz, Phys. Rev. B {\bf 12}, 5007 (1975).

\bibitem{CsCoCl3}
M. Mekata and K. Adachi, J. Phys. Soc. Jpn. \textbf{44}, 806 (1978).

\bibitem{visser}
D. Visser, G. C. Verschoor, and D. J. W. Ijdo, 
Acta Crystallogr. B\textbf{36}, 28 (1980).

\bibitem{ex55}
K. Adachi, K. Takeda, F. Matsubara, M. Mekata and T.~Haseda, 
J. Phys. Soc. Jpn. \textbf{52}, 2202 (1983).

\bibitem{RbFeBr3-experi}
T. Mitsui, K. Michida, T. Kato, and K. Iio,
J. Phys. Soc. Jpn. {\bf 63}, 839 (1994).

\bibitem{morishita52}
K. Morishita, T. Kato, K. Iio, T. Mitsui, M. Nasui, T. Tojo and T. Atake,
Ferroelectrics \textbf{238}, 105 (2000).

\bibitem{morishitaC}
K. Morishita, K. Iio, T. Mitsui and T. Kato,
J. Magn. Magn. Mater. {\bf 226-230}, 579 (2001).

\bibitem{nishiwaki2}
Y. Nishiwaki, T. Kato, Y. Oohara and K. Iio,
J. Phys. Soc. Jpn. {\bf 73}, 2841 (2004).

\bibitem{experi-mag}
Y. Nishiwaki, H. Imamura, T. Mitsui, H. Tanaka and K. Iio,
J. Phys. Soc. Jpn. {\bf 75}, 094702 (2006).

\bibitem{experi-f}
Y. Nishiwaki, A. Oosawa, T. Nakamura, K. Kakurai, N. Todoroki,
N. Igawa, Y. Ishii, and T. Kato, to appear in J. Phys. Soc. Jpn.

\bibitem{shiba}
H. Shiba, Prog. Theor. Phys. \textbf{64}, 466 (1980).

\bibitem{matsubara-ina}
F. Matsubara and S. Inawashiro, J. Phys. Soc. Jpn. \textbf{53}, 4373 (1984).

\bibitem{kurata}
T. Kurata and H. Kawamura, J. Phys. Soc. Jpn. \textbf{64}, 232 (1995).

\bibitem{koseki}
O. Koseki and F. Matsubara, J. Phys. Soc. Jpn. \textbf{69}, 1202 (2000).

\bibitem{todoroki}
N. Todoroki and S. Miyashita, J. Phys. Soc. Jpn. \textbf{73}, 412 (2004).

\bibitem{shiramaster}
T. Shirahata and T. Nakamura, J.  Phys. Soc. Jpn. {\bf 73}, 254 (2004).

\bibitem{nishi-todo.MF}
Y. Nishiwaki and N. Todoroki, J.  Phys. Soc. Jpn. {\bf 75}, 024708 (2006).



\bibitem{ashkin}
J. Ashkin and E. Teller, Phys. Rev. {\bf 64}, 178 (1943);
C. Fan, Phys. Lett. {\bf 39A}, 136 (1972).

\bibitem{kawamura-spin-lattice}
M. L. Plumer, A. Caill\'e and H. Kawamura,
Phys. Rev. B {\bf 44}, 4461 (1991).

\bibitem{axcMC}
T. Nakamura, in preparation.

\bibitem{chb}
O. Koseki and F. Matsubara, J. Phys. Soc. Jpn. {\bf 66}, 322 (1997);
F. Matsubara, A. Sato, O. Koseki and T. Shirakura, Phys. Rev. Lett. 
{\bf 78}, 3237 (1997).

\bibitem{meloche}
E. Meloche and M. L. Plumer,
Phys. Rev. B {\bf 76}, 174430 (2007).


\bibitem{totaqmc}
T. Nakamura and Y. Ito, J. Phys. Soc. Jpn. {\bf 72}, 2405 (2003).

\bibitem{suzukitrotter}
{\it Quantum Monte Carlo Methods in Condensed Matter Physics}, 
ed. M. Suzuki (World Scientific, Singapore, 1994).

\bibitem{mix-state-start1}
C. Rebbi, Phys. Rev. D {\bf 21}, 3350 (1980). 

\bibitem{mix-state-start2}
F. Fucito and A. Vulpiani, Phys. Lett. A {\bf 89}, 33 (1982).

\bibitem{mix-state-start3}
Y. Ozeki, K. Kasono, N. Ito and S. Miyashita,
Physica A {\bf 321}, 271 (2003).




\bibitem{miyashita-spincrossover}
S. Miyashita {\it et al.}, Phys. Rev. B {\bf 77}, 014105 (2008).

\bibitem{kakurai}
N. Ikeda {\it et al.}, Nature(London) {\bf 436}, 1136 (2005).

\bibitem{ishihara}
A. Nagano, M. Naka, J. Nasu, and S. Ishihara,
Phys. Rev. Lett. {\bf 99}, 217202 (2007).

\end{thebibliography}
\end{document}